\documentclass[reprint,amsmath,twocolumn,amssymb,aps,pra,superscriptaddress,tightenlines]{revtex4}
\usepackage{graphicx}
\usepackage{bm}
\usepackage{ulem}
\usepackage{color}
\usepackage{mathrsfs}
\usepackage{slashed}
\allowdisplaybreaks[4]
\usepackage[colorlinks,citecolor=blue,urlcolor=blue,linkcolor=blue]{hyperref}
\usepackage{appendix}
\begin{document}

	\preprint{APS/123-QED}
	
	\title{Parity-time-symmetric two-qubit system: entanglement and sensing}
	\author{Ji Zhang}\thanks{Co-first authors with equal contribution}
	\affiliation{Key Laboratory of Low-Dimension Quantum Structures and Quantum Control of Ministry of Education, Synergetic Innovation Center for Quantum Effects and Applications, and Department of Physics, Hunan Normal University, Changsha 410081, China}
	\affiliation{Interdisciplinary Center for Quantum Information, National University of Defense Technology, Changsha 410073, China}
	\author{Yan-Li Zhou}\thanks{Co-first authors with equal contribution}
	\affiliation{Interdisciplinary Center for Quantum Information, National University of Defense Technology, Changsha 410073, China}
	\affiliation{Institute for Quantum Science and Technology, College of Science, National University of Defense Technology, Changsha  410073, China}
	\author{Yun-lan Zuo}
	\affiliation{Key Laboratory of Low-Dimension Quantum Structures and Quantum Control of Ministry of Education, Synergetic Innovation Center for Quantum Effects and Applications, and Department of Physics, Hunan Normal University, Changsha 410081, China}
    \author{Ping-Xing Chen}\email{pxchen@nudt.edu.cn}
	\affiliation{Interdisciplinary Center for Quantum Information, National University of Defense Technology, Changsha 410073, China}
	\affiliation{Institute for Quantum Science and Technology, College of Science, National University of Defense Technology, Changsha  410073, China}
	 \author{Hui Jing}\email{jinghui@hunnu.edu.cn}
	\affiliation{Key Laboratory of Low-Dimension Quantum Structures and Quantum Control of Ministry of Education, Synergetic Innovation Center for Quantum Effects and Applications, and Department of Physics, Hunan Normal University, Changsha 410081, China}
    \affiliation{Synergetic Innovation Academy for Quantum Science and Technology, Zhengzhou University of Light Industry, Zhengzhou 450002, China}
    \author{Le-Man Kuang}\email{lmkuang@hunnu.edu.cn}
	\affiliation{Key Laboratory of Low-Dimension Quantum Structures and Quantum Control of Ministry of Education, Synergetic Innovation Center for Quantum Effects and Applications, and Department of Physics, Hunan Normal University, Changsha 410081, China}
    \affiliation{Synergetic Innovation Academy for Quantum Science and Technology, Zhengzhou University of Light Industry, Zhengzhou 450002, China}

 \date{\today}

\begin{abstract}
In this paper we study exceptional-point (EP) effects and  quantum sensing in a parity-time (PT)-symmetric two-qubit system with the Ising-type interaction. We explore EP properties of the system by analyzing degeneracy of energy eigenvalues or entanglement of eigenstates. We investigate entanglement dynamics of the two qubits in detail. In particular,  we demonstrate that
the system can create the steady-state entanglement in the PT-broken phase and collapse-revival phenomenon of entanglement in the PT-symmetric phase during the long-time evolution.  We show that entanglement can be generated more quickly  than the corresponding Hermitian system. Finally, we prove that the sensitivity of eigenstate quantum sensing for the parameters exhibits the remarkable enharncement at EPs, and propose a quantum-coherence measurement to witness the existence of EPs.
\end{abstract}
\maketitle
	
\section{Introduction}

Symmetry is the most fundamental property of physical systems and the root cause of many physical phenomena such as the law of conservation of mechanical quantities. In the year 1998, Bender and Boettcher \cite{1} found that a non-Hermitian Hamiltonian with PT symmetry could produce purely real spectra. It made an important implication for the later development of non-Hermitian quantum mechanics which beyond conventional quantum mechanics.  Mostafazadeh \cite{2,3} further introduced and explored the pseudo-Hermitian operators with a more general sense of PT symmetry. The concept of PT symmetry has gone far beyond the scope of quantum mechanics and has penetrated many branches of physics. Ruschhaupt and coworkers \cite{4} showed that the optical system with PT symmetry can be constructed by designing the refractive index of the optical medium. El-Ganainy et al. \cite{5} proposed the concept of waveguide optics with PT symmetry, and showed that the optical system could provide a simple but not mediocre ideal physics system for the study of PT symmetry physics. Near-axis PT symmetric optics was further developed by Makris et al \cite{6,7}. Subsequently, the theory of PT-symmetric optics was confirmed by many experiments \cite{8,9,10,11,12}. These works motivate a new tidal wave in PT-symmetric optics and photonics \cite{13,14,15,16,17,18,19,20,21}.

Quantum systems with  PT symmetry are important not only to studying quantum mechanics but also providing the theoretical foundation to many other branches of physics \cite{22,23,24,25,26,27,28,29,30,31}. In a PT-symmetric quantum system, when by continuously tuning a parameter the energy spectrum changes from real into complex conjugate pairs, and the corresponding eigenvectors no longer exhibit the underlying PT-symmetry. This transition from PT-symmetric (PTS) to PT-broken (PTB) phase occurs at an exceptional point (EP) with the order of  $n$ (EP$n$), where both $n$ eigenvalues and eigenvectors of the system coalesce \cite{M.-A. Miri,F. Minganti}. As a result, many counterintuitive phenomena emerge in such systems, for example, a change of the nature of dynamics from oscillations to the nonunitary time evolution occurs at the EP, which have been experimentally  observed in a variety of physical systems, such as in photonic waveguides \cite{K,C}, optical microcavities \cite{Weijian Chen,Steffen Richter}, coupled atom-cavity systems \cite{Youngwoon Choi}, mechanical system \cite{Bender}, resonators \cite{L,H}, and also in quantum realm \cite{Y. Wu,M. Naghiloo,W. C. Wang}.

In this paper, we study EP effects and  quantum sensing in a PT-symmetric two-qubit system with the Ising-type interaction. We find that inter-qubit entanglement exhibits  very different dynamic behaviors  in the PTS and PTB phases.  We show that EP characteristics of the system can be described through the degeneracy of energy eigenvalues or entanglement of eigenstates.
We find that the collapse-revival (CR) phenomenon of entanglement appears in the PTS phase while steady-state entanglement can be generated in the PTB phase during the long-time dynamic evolution.  We also find that entanglement can be created more quickly in the PT-symmetric system than in the corresponding Hermitian system.  We investigate EP quantum sensing by calculating the quantum Fisher information (QFI) to estimate the parameters of the system and propose a quantum-coherence measurement scheme to witness the PTS-PTB phase transition.
	
The rest of this paper is organized as follows. In Sec. II, we describe a parity-time symmetric two-qubit system with the Ising-type interaction and analyze EP properties of the system.  In sec. III, we investigate inter-qubit entanglement dynamics to reveal EP effects of the system including the steady-state entanglement, the entanglement CR effect, and rapid generation of entanglement.  In sec. IV, we study EP quantum sensing via the QFI and propose the witnessing scheme of the PTS-PTB phase transition. In Sec. V , We discuss experimental feasibility of the PT-symmetric two-qubit system.  The last section is devoted to conclusions and a brief discussion.

\begin{figure}[b]
		\includegraphics[height=0.15\textheight]{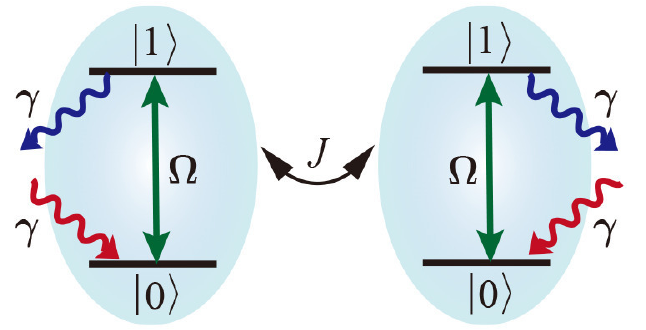}
		\caption{ Schematic diagram of our model which consists of two identical PT symmetric two-level systems. Here J denotes the interaction between two PT symmetric systems.} \label{fig1}
	\end{figure}

\section{Model and  exceptional points}

We consider a non-Hermitian system shown in Fig. 1, which consists of a  PT symmetric two-qubit system with the same loss and gain, and the Ising-type interaction. The relevant internal level structure for each qubit is given by the ground state $|0\rangle$ and the excited state $|1\rangle$. Assuming that the two qubits under consideration are
identical, our model is well described by the non-Hermitian Hamiltonian
	
\begin{align}
		H&=H^{1}_{PT}+ H^{2}_{PT}+J{\sigma^{1}_{z}}\otimes \sigma^{2}_{z} \label{PT coupling}.
	\end{align}
where $H^i_{PT}=(\Omega\sigma^i_{x}-i\gamma\sigma^i_z)/2$ is the PT-symmetric Hamiltonian of each qubit with balanced gain and loss $\gamma$ under the resonant condition, the coupling rate between two internal states of each qubit is $\Omega$, $J$ denotes the coupling strength between the two qubits, and $\sigma_{(x,z)}$ is a Pauli matrix. The parity operator of the two-qubit system can be expressed as $P=\sigma^{1}_{x}\otimes \sigma^{2}_{x}$ and the time reversal operator $T$ is the complex conjugation operation.  The Hamiltonian (1) is a PT-symmetric Hamiltonian due to $[PT, H]=0$.

It is straightforward to  solve the Hamiltonian $H $ given by Eq. (1) with the following four eigenvalues
	\begin{align}
		E_{1}=&-J,   \hspace{0.3cm} E_{2}=\frac{1}{3}\left(J+XY^{-1}+Y\right),  \\
		E_{3}=&\frac{1}{3}\left(J-XY^{-1}e^{i\frac{\pi}{3}}-Ye^{-i\frac{\pi}{3}}\right), \\
        E_{4}=&\frac{1}{3}\left(J-XY^{-1}e^{-i\frac{\pi}{3}}-Ye^{i\frac{\pi}{3}}\right), \label{eigenvalues 1}
	\end{align}
where we have introduced three parameters
	\begin{align}
		X=&4J^2+3\Omega^2-3\gamma^2,  \\
		Y=&\sqrt[3]{-8J^3-9J(\Omega^2+2\gamma^2)+3\sqrt{3}\sqrt{Z}}, \\
		Z=&16J^4\gamma^2+J^2(8\gamma^4+20\gamma^2\Omega^2-\Omega^4)+(\gamma^2-\Omega^2)^3. \label{eigenvalue}
	\end{align}

We now investigate EP properties of the PT-symmetric Hamiltonian (1).  An analytic EP analysis of Hamiltonian (1) is given in Appendix A where it is indicated that there exists the second-order EPs of the  Hamiltonian (1) for two energy eigenvalues $E_3$  and $E_4$.  These EPs are described by two critic parameters $(J_c, \Omega_c)$, they construct an EP curve line which is determined by Eq. (A8) in Appendix A.  At the EP curve line, both of two energy eigenvalues ($E_3$ and $E_4)$ and their corresponding eigenstates are degenerate, at the same time the imaginary parts of two energy eigenvalues vanish.
In Fig. 2, the real and imaginary parts of the degenerate eigenvalues $E_3$ and $E_4$ are plotted with respect to two parameters $(J, \Omega)$, respectively.  The eigenvalue is marked by green (yellow) for $E_3 ( E_4)$. The green-yellow intersection forms the EP curve.

\begin{figure*}[t]
\includegraphics[width=0.9\textwidth]{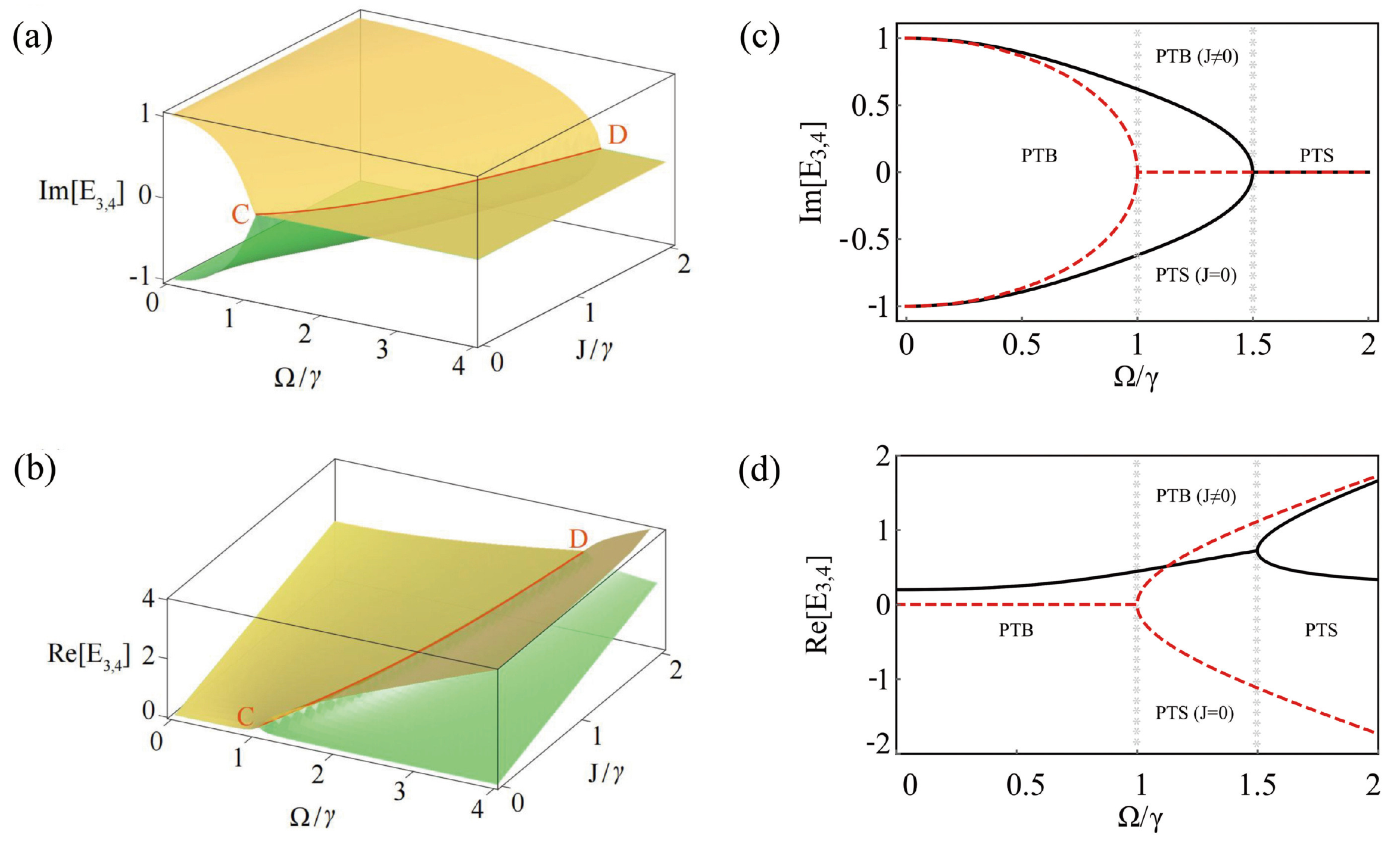}
\caption{The real and imaginary parts of  eigenvalues  $E_3$  and $E_4$. The intersection of  $E_3$  and $E_4$ is  the EP curve. Here and in other figures below, all plotted quantities are dimensionless, i.e., $\gamma=1.0$.} \label{fig2}
\end{figure*}

For the PT-symmetric Hamiltonian given by Eq. (1), the PT symmetry of its eigenstates can be spontaneously broken due to the appearance of the nonzero imaginary part of the eigenvalue $E_i$. The PT-broken region sensitively depends on parameters $\Omega$ and $\gamma$. In Fig. 2, the dashed curve CD is the EP cureve with the  critic parameters $(J_c, \Omega_c)$.
From Fig. 2 (b) we can see that the  PT-broken region ($\Omega<\Omega_c$) is on the left-hand-side regime of the EP curve,  while the  PT-unbroken region ($\Omega>\Omega_c$) is on the right-hand-side regime of the EP curve.

The corresponding eigenvectors are given by
	\begin{align}
		|\Psi_{1}\rangle=&\frac{1}{\sqrt{2}}(|10\rangle - |01\rangle),  \\
		|\Psi_{2}\rangle=&N_{2}(R_{21}|00\rangle+R_{22}|01\rangle+R_{22}|10\rangle+|11\rangle),  \\
		|\Psi_{3}\rangle=&N_{3}(R_{31}|00\rangle+R_{32}|01\rangle+R_{32}|10\rangle+|11\rangle),   \\
		|\Psi_{4}\rangle=&N_{4}(R_{41}|00\rangle+R_{42}|01\rangle+R_{42}|10\rangle+|11\rangle),\label{eigenvalues 2}
	\end{align}
where the relevant coefficients are given by
\begin{align}
       	R_{j1}=&-2(J+E_{j})(J-E_{j}+i\gamma)/\Omega^{2}-1,  \\
        R_{j2}=&-(J-E_{j}+i\gamma)/\Omega,  \\
       	N_{j}=&\left(1 + |R_{j1}|^2 + 2|R_{j2}|^2\right)^{-1/2},  \hspace{0.2cm} (j=2,3,4).
 \end{align}

It is interesting to note that the EP curve can be characterized not only by the real and imaginary parts of  eigenvalues of the two-coupled-qubit system  but also by calculating the quantum concurrence of eigenstates. The concurrence $C$ of two qubits with a density operator $\hat{\rho} $ is given by \cite{William:1998,X. F,Z. H.}
\begin{gather}
C=\text{Max}\{0,\sqrt{\lambda_1}-\sqrt{\lambda_2}-\sqrt{\lambda_3}-\sqrt{\lambda_4}\}, \label{concurrency}
\end{gather}
where the numbers $\lambda_i$ ($i=1,2,3,4$) are the square roots of the eigenvalues in nonincreasing order of the matrix $\tilde{\rho}$,
\begin{align*}
       	\tilde{\rho}=\rho (\sigma^1_y\otimes\sigma^1_y) \rho^* (\sigma^1_y\otimes\sigma^2_y),
      \end{align*}
where $\rho^*$ is the complex conjugation of $\rho$, and $\sigma_y$ is Pauli matrix in the computational basis.

In the what follows,  we study   entanglement properties of two  eigenstates  $|\Psi_{3}\rangle$ and $|\Psi_{4}\rangle$ in the PTS and PTB phases.

\begin{figure}[h]
		\includegraphics[width=0.4\textwidth]{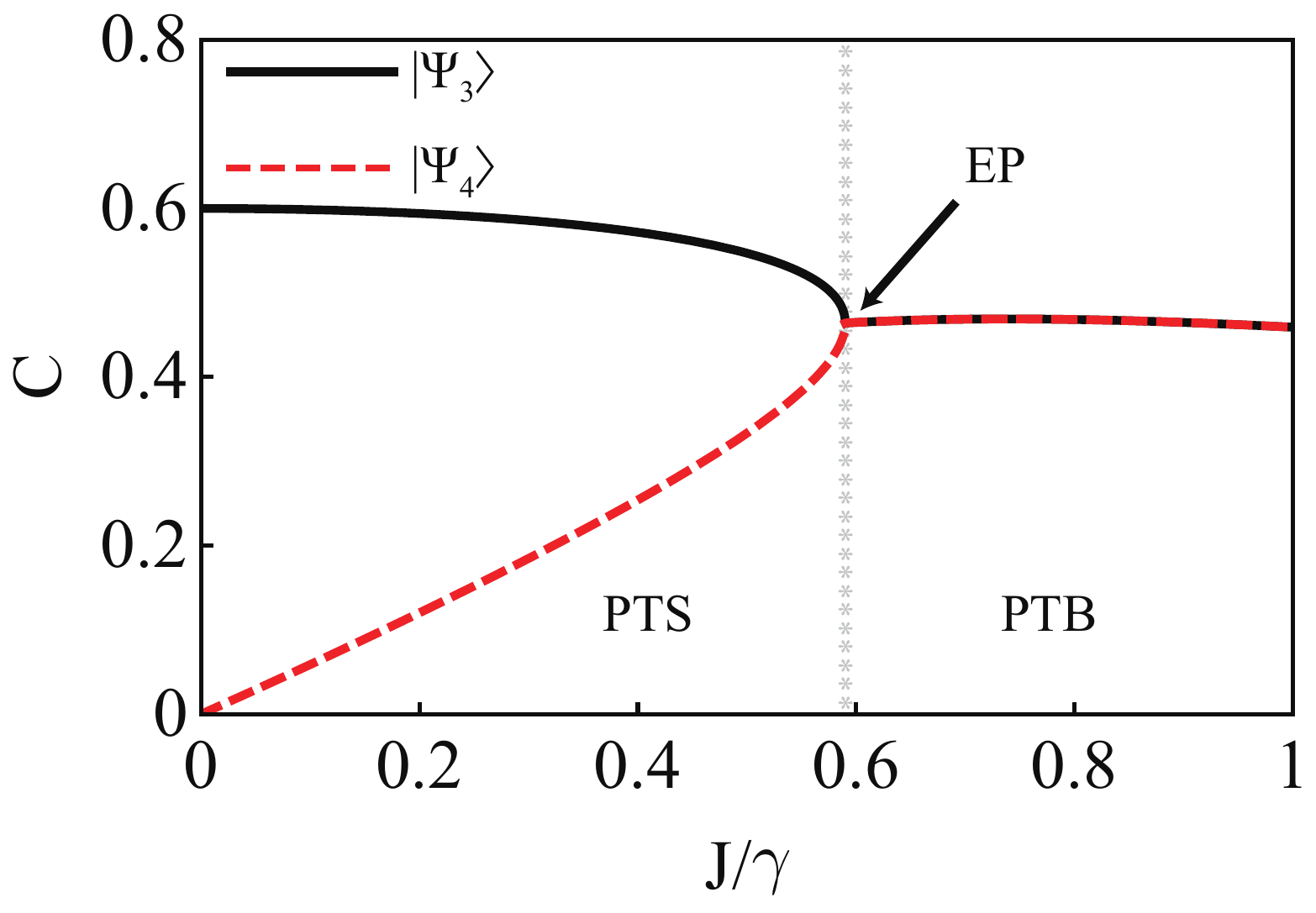}
\put(-215,135){(a)}

\includegraphics[width=0.4\textwidth]{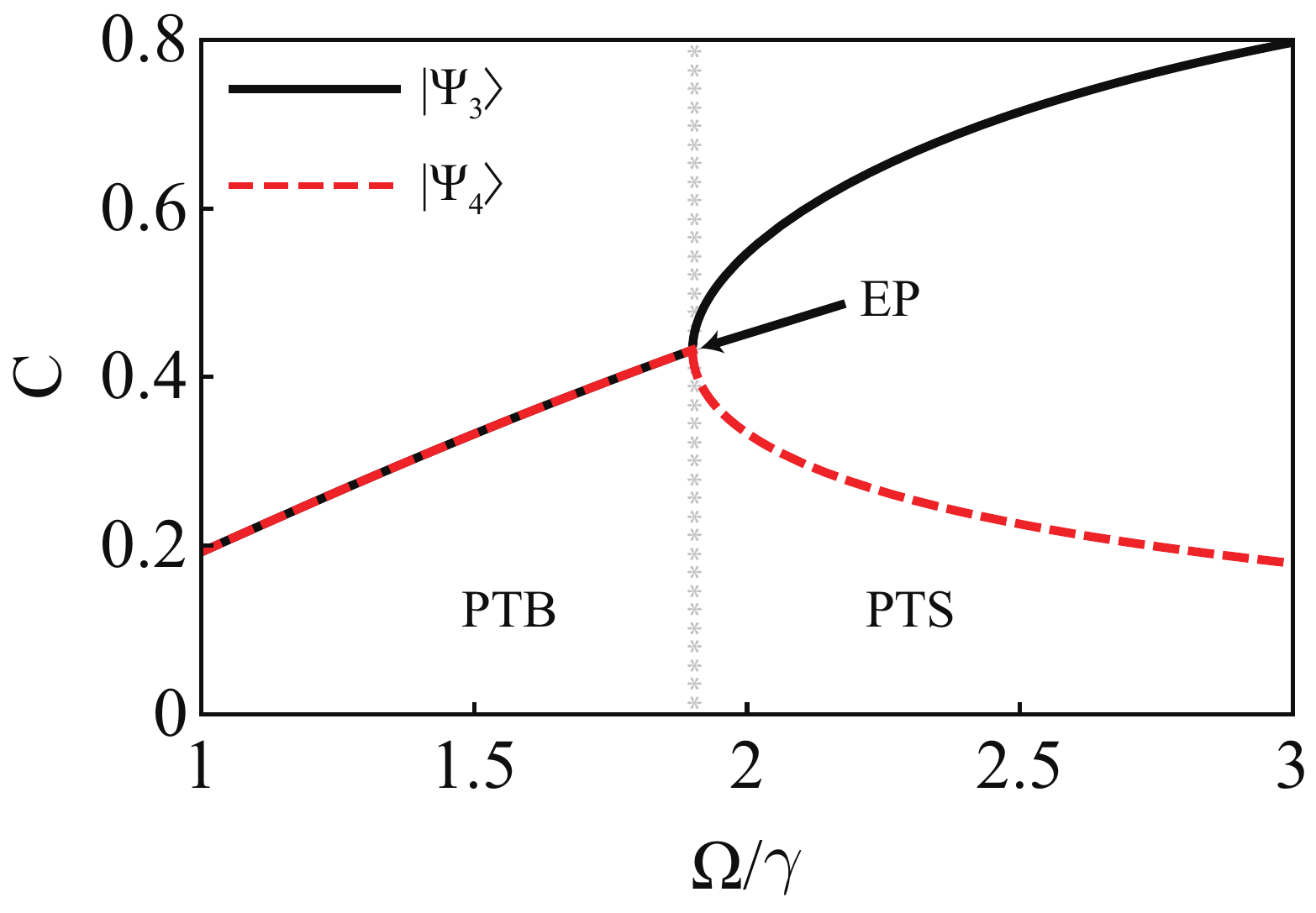}
\put(-215,135){(b)}
\caption{ The concurrence of   eigenstates $|\Psi_{3}\rangle$ and $|\Psi_{4}\rangle$   with respect to the system parameters $J$ and $\Omega$. Here the solid and dashed lines correspond to $|\Psi_{3}\rangle$ and $|\Psi_{4}\rangle$, respectively. (a) The concurrence of  $|\Psi_{3}\rangle$ and $|\Psi_{4}\rangle$ varies with the parameter $J$  with the critic value  $J_c=0.588$ for the fixed value of $ \Omega=2.000$.  (b) The concurrence of  $|\Psi_{3}\rangle$ and $|\Psi_{4}\rangle$ varies with the parameter $\Omega$ with the critic value  $ \Omega_c=1.649$  for the fixed value of $J=0.300$.} \label{figc}
\end{figure}

Their corresponding density matrixs $\tilde{\rho}_{s} $ can be easily derived and written as
\begin{equation}
 \tilde{\rho}_{s}=|N_{s}|^4\begin{pmatrix}
  a R_{s1} & bR_{s1} R^{\ast}_{s2} &bR_{s1} R^{\ast}_{s2}  & |R_{s1}|^{2}a\\
 aR_{s2}  &b|R_{s2}|^{2}&b|R_{s2}|^{2}& R^{\ast}_{s1}R_{s2}a \\
 aR_{s2}  &b|R_{s2}|^{2}&b|R_{s2}|^{2} &R^{\ast}_{s1}R_{s2}a \\
 a  & bR^{\ast}_{s2} & bR^{\ast}_{s2} & R^{\ast}_{s1}a
\end{pmatrix}.  \label{Hamiltonian}
    \end{equation}
where $N_{s}$ $(s=3,4)$ is given by Eq. (14), and $a=R_{s1}+R_{s1}^{\ast}-2|R_{s2}|^{2},b=2[|R_{s2}|^{2}- \text{Re}(R_{s1})]$. The four roots of the eigenvalues of the matrixs $\tilde{\rho}_{s}$ are
\begin{align}
\lambda_{s1}=&1/2|N_{s}|^{4}[(R_{s1}+R_{s1}^{\ast})^{2}\notag \\
& +a^{2}+2b|R_{s2}|^{2}-4|R_{s2}|^{2}\text{Re}(R_{s1})], \notag \\
\lambda_{s2}=&1/2|N_{s}|^{4}[(R_{s1}+R_{s1}^{\ast})^{2}\notag \\
&-a^{2}+2b|R_{s2}|^{2}-4|R_{s2}|^{2}\text{Re}(R_{s1})], \notag \\
\lambda_{s3}=& \lambda_{s4}=0. \label{PTB}
\end{align}
Then we  can obtain the concurrence $C$  of  eigenstates $|\Psi_{s}\rangle$ with the following expression
\begin{align}
C(\rho_{s})=&\sqrt{\lambda_{s1}}-\sqrt{\lambda_{s2}}, \hspace{0.5cm} (s=3, 4).
\end{align}

We numerically investigate EPs characteristics from the point of view of quantum entanglement of eigenstates of the system Hamiltonian.   In Fig. 3, we plot the concurrence of   eigenstates $|\Psi_{3}\rangle$ and $|\Psi_{4}\rangle$   with respect to the system parameters $J$ and $\Omega$. Here the solid and dashed lines correspond to $|\Psi_{3}\rangle$ and $|\Psi_{4}\rangle$, respectively. Firstly,  we fix the parameter $\Omega$ and take $J$ as the variable.   Fig. 3(a) shows that the concurrence of  $|\Psi_{3}\rangle$ and $|\Psi_{4}\rangle$ varies with the parameter $J$ for the fixed value of $ \Omega=2.000$.  The EP appears at the critical value  $J_c=0.588$.  From Fig. 3(a) we can see that the  PTS region ($J<J_c$) is on the left-hand-side regime of the EP  while the  PTB region ($J>J_c$) is on the right-hand-side regime of the EP. In the PTB region, two eigenstates $|\Psi_{3}\rangle$ and $|\Psi_{4}\rangle$ have the same concurrence which almost remains unchange with the increase of $J$. In the PTS region, two eigenstates $|\Psi_{3}\rangle$ and $|\Psi_{4}\rangle$ have different concurrence. And the concurrence of $|\Psi_{3}\rangle$ ($|\Psi_{4}\rangle$) decreases (increases) with the increase of $J$. They reach the same concurrence when $J$ approaches $J_c$ (the EP).

However, the situation would be very different when we fix the parameter $J$ and take $\Omega$ as the variable.
Fig. 3(b) indicates that the concurrence of  $|\Psi_{3}\rangle$ and $|\Psi_{4}\rangle$ varies with the parameter $\Omega$ for the fixed value of $J=0.300$.  The EP appears at the critical value  $\Omega_c=1.649$.  From Fig. 3(b) we can see that the  PTB region ($\Omega<\Omega_c$) is on the left-hand-side regime of the EP  while the  PTS region ($\Omega>\Omega_c$) is on the right-hand-side regime of the EP. In the PTS region, two eigenstates $|\Psi_{3}\rangle$ and $|\Psi_{4}\rangle$ have differen concurrence.  And the concurrence of $|\Psi_{3}\rangle$ ($|\Psi_{4}\rangle$) decreases (increases) with the decrease of $\Omega$. They reach the same concurrence when $\Omega$ approaches $\Omega_c$ (the EP). In the PTB region, two eigenstates $|\Psi_{3}\rangle$ and $|\Psi_{4}\rangle$ have the same concurrence which increases with the increase of $\Omega$.
Above analytic and numerical studies indicate that the PT-symmetric two-qubit system have a second-order EP curve, and relevant eigenstates exhibit different entanglement characteristics in the PTS and PTB regions.

\section{EP effects on entanglement dynamics}
 Quantum entanglement is a counterintuitive nonlocal correlation, which plays an
extremely important role in the study of fundamental problems in quantum mechanics \cite{Valerio:2009,Peter:2000,Lo:2005,Dik:1997,Riebe:2004,Bose:2000,ZZLi,Jiao1,Jiao2,BLi,Kuang1,Kuang2}. It is also widely used in quantum information processing \cite{Horodecki,Nielsen,Henderson:2000}. In this  section, we investigate the influence of EPs on quantum  entanglement between two qubits in dynamic evolution. We will reveal three types of dynamic EP effects including the steady-state entanglement, the entanglement CR effect, and rapid generation of entanglement.

\subsection{Steady-state entanglement and CR phenomenon}

We now numerically explore the properties of entanglement dynamics in the PT-symmetric two-qubit system under our consideration.  We will pay the main attention to dynamics of entanglement in the long-time evolution to reveal steady-state entanglement and the EP-induced CR phenomenon of entanglement.

\begin{figure}[b]
\includegraphics[width=0.42\textwidth]{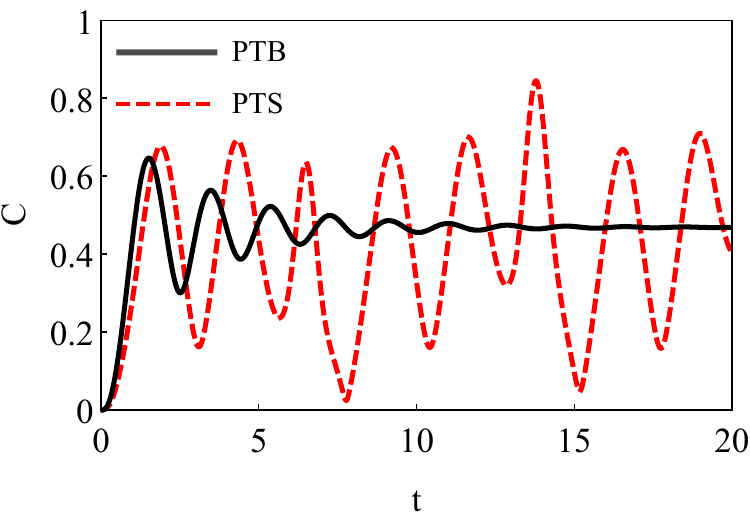}
\caption{The dynamic evolution of entanglement described by the concurrence in the PTS and PTB phases. Here the dashed and solod lines correspond to the PTS and PTB phases, respectively.} \label{fig2}
\end{figure}

Assuming that the two-qubit system under our consideration is initially the unentangled state $|0, 0\rangle$, we can study numerically quantum dynamics of the system with the Hamitonian (1). In Fig. 4, we plot the concurrence to describe the dynamic evolution of entanglement in the PTS and PTB phases. Here the dashed and solid lines correspond to the PTS and PTB phases, respectively.   The relevant parameters are taken as $(\Omega, J)=(2.0, 0.7)$ and $(2.0, 0.4)$  for the PTB and PTS phases while the critical values of parameters at the EP are $(\Omega_c, J_c)=(2.0, 0.6)$.

 From Fig. 4 we can observe that the PT-symmetric two-qubit system has very different dynamic behaviors of entanglement in the PTS and PTB phases.  In the PTS phase,  entanglement dynamics exhibit continuous oscillations in the whole dynamic evolution. In contrast, in the PTB phase,  entanglement dynamics decay rapidly in a short time during the time evolution to a steady state with a fixed value of the concurrence. The amount of steady-state entanglement can reach $C\approx 0.5$. Therefore, the phase transition from the PTS phase to the PTB  phase can induce the entanglement transition of the system from oscillating entanglement to steady-state entanglement. It is worthwhile to mention that various quantum applications such as quantum communication \cite{55,56,57}, quantum computation \cite{58}, quantum metrology \cite{59}, and quantum sensing \cite{60} often depend on the ability to generate long-lived or steady-state entanglement. The PTS-PTB transition in the PT-symmetric system provides a possible way to generate steady-state entanglement.

In order to further observe the long-time behaviors of entanglement dynamics,  in Fig. 5  we plot the long-time evolution of entanglement in the PTS phase with respect to two parameters $J$  and $\Omega$ when the system in initially in the state $|00\rangle$.
From Fig. 5, it is particularly interesting to observe that entanglement oscillations in the PTS phase exhibit the phenomenon of collapses and revivals during the long-time evolution.
The CR phenomenon is sensitive to the change of the parameters $J$ and $\Omega$.
Obviously, the system parameters  $J$ and $\Omega$ have an important influence on the entanglement CR.
Increasing (decreasing) the parameter $J$  ($\Omega$) slightly from $J=0.336$  ($\Omega=1.902$)  to $J=0.337$  ($\Omega=1.901$), the revival time is seen to increase by about two times in Fig. 5.
We note that the critic parameters of the EPs in Fig. 5(a) and 5(b) are $(\Omega_c, J_c) =(1.700,0.338)$  and $(J_c, \Omega_c)=(0.500,1.900)$, respectively. Fig. 5 reflects the fact that the closer to the critical values at EPs the parameters   $J$  and $\Omega$ are, the longer the revival time becomes.
Hence we can effectively manipulate the revival time through tuning the parameters $J$ and $\Omega$.


\begin{figure}[h]
		\includegraphics[width=0.42\textwidth]{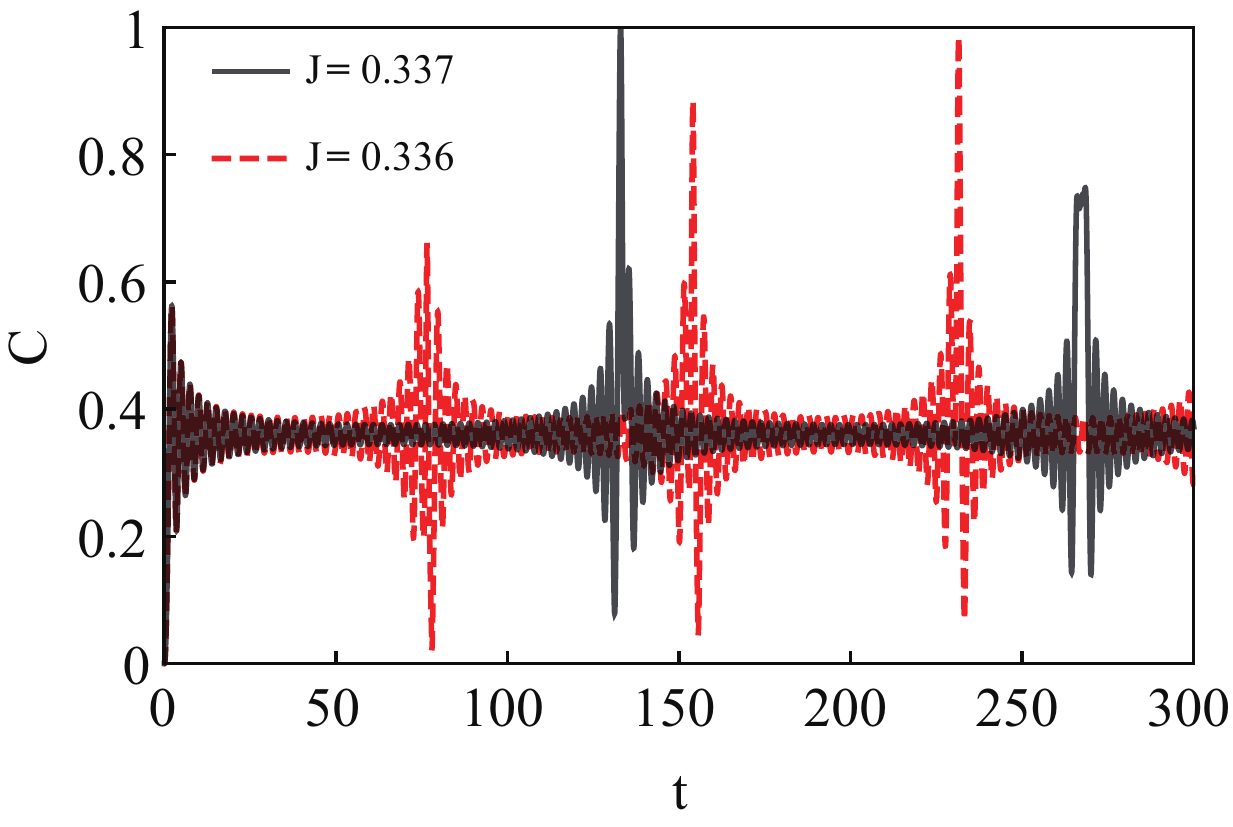}
\put(-215,135){(a)}

\includegraphics[width=0.42\textwidth]{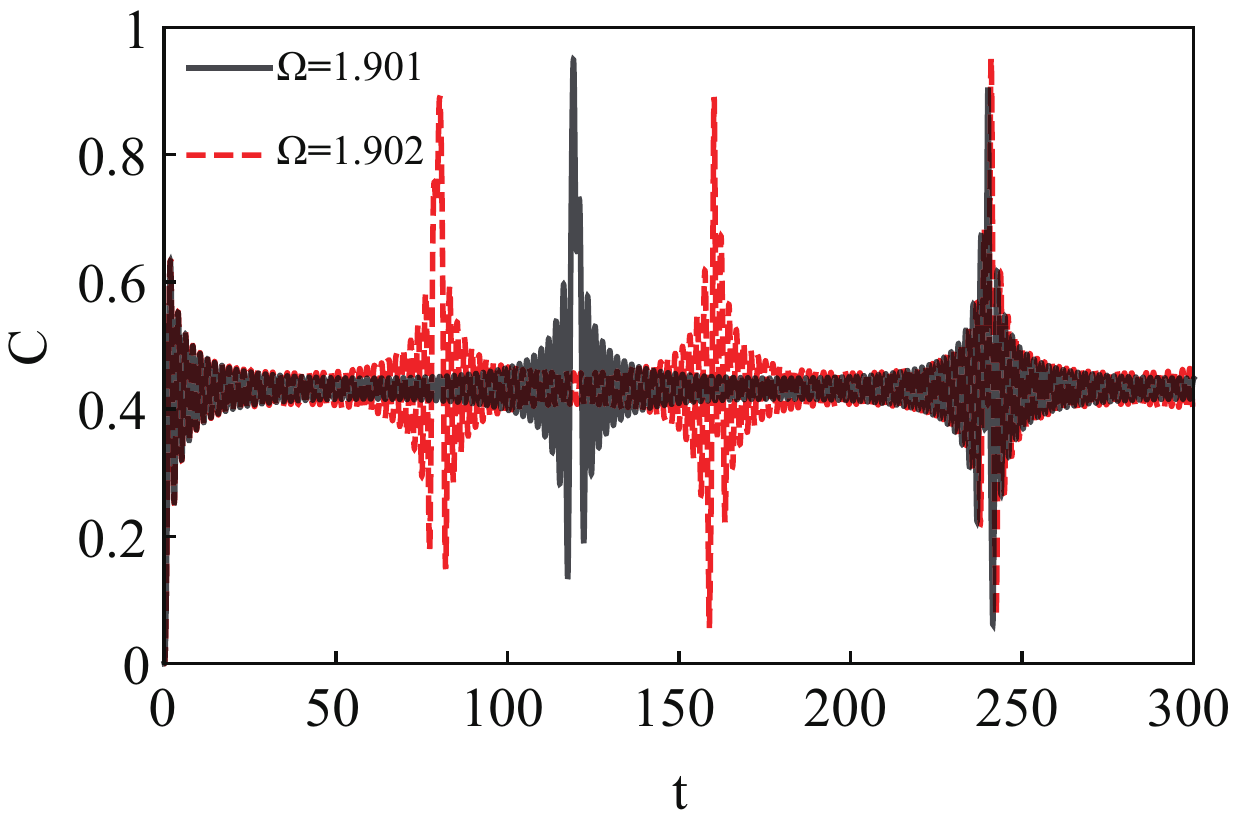}
\put(-215,135){(b)}
		\caption{The long-time dynamic  evolution of entanglement described by the concurrence in the PTS  phase. (a)The revival time manipulation via changing the inter-qubit interaction. The  parameter takes $\Omega=\Omega_c$ with  the  critic parameters being $(\Omega_c, J_c) =(1.700,0.338)$. The black solid line (red dashed line) parameters are $\Omega=1.700$, $J=0.337$ ($\Omega=1.700$, $J=0.336$). (b) The revival time manipulation via changing the qubit parameter.  The  parameter takes $J=J_c$ with  the  critic parameters being $(J_c, \Omega_c)=(0.500,1.900)$. The black solid line (red dashed line) corresponds to $\Omega=1.901$, $J=0.500$ ($\Omega=1.902$, $J=0.500$). The initial state of the system is direct product state $\Psi_0=|00\rangle$.} \label{fig11}
\end{figure}

\subsection{Rapid generation of entanglement}
In the following, we analyze how the exceptional point accelerates generation of entanglement in the dynamic evolution. We suppose that the PT symmetric system under consideration is initially in a coherent superposition state and the other is in the ground state. Thus, the total initial state of our system is given by
\begin{equation}
 |\Psi_0\rangle=(\rm sin\theta|0\rangle+\rm cos\theta|1\rangle)\otimes|0\rangle.
\end{equation}

Firstly, we consider the Hermitian case with  $\gamma = 0$. Fig. 6(a) represents the dynamic evolution of entanglement between two qubits when the parameters are taken as $\gamma=0$, $J=0.01$, and $\Omega=1.5$. From  Fig. 6(a) we can see that the entanglement evolution is
periodic. And we can find that the quantum coherence of the initial state significantly affects the generated maximal entanglement and the evolution period. The dynamic evolution process can create maximum entanglement with $\mathcal{C}= 1$ when quantum coherence of the first qubit vanishes ( $\theta=\pi/2$) while the maximal amount of entanglement generated in the dynamic evolution achieves only $\mathcal{C} \approx 0.5$  when the first qubit has maximal quantum coherence ($\theta = \pi/4$).

Then we take account into the non-Hermitian case with $\gamma \neq 0$. In Fig. 6(b) we plot the dynamic evolution of entanglement between two qubits in the PTS phase of the PT-symmetric system when the parameters are taken as  $\gamma=1.1$, $J=0.01$, and $\Omega=1.5$.  From Fig. 6(b) we can observe a remarkable phenomenon that the maximal entanglement with $C=1$ can be produced in a much short time in the PTS phase.  Comparing Fig. 6(a) and Fig. 6(b), we can find that the time required for the non-Hermitian system to produce maximum entanglement is significantly shorter than that of the Hermitian system. Therefore, non-Hermitian dynamics could accelerate the production of quantum entanglement, which would lead to rapid preparation of entanglement.

\begin{figure}[t]
		\includegraphics[width=0.42\textwidth]{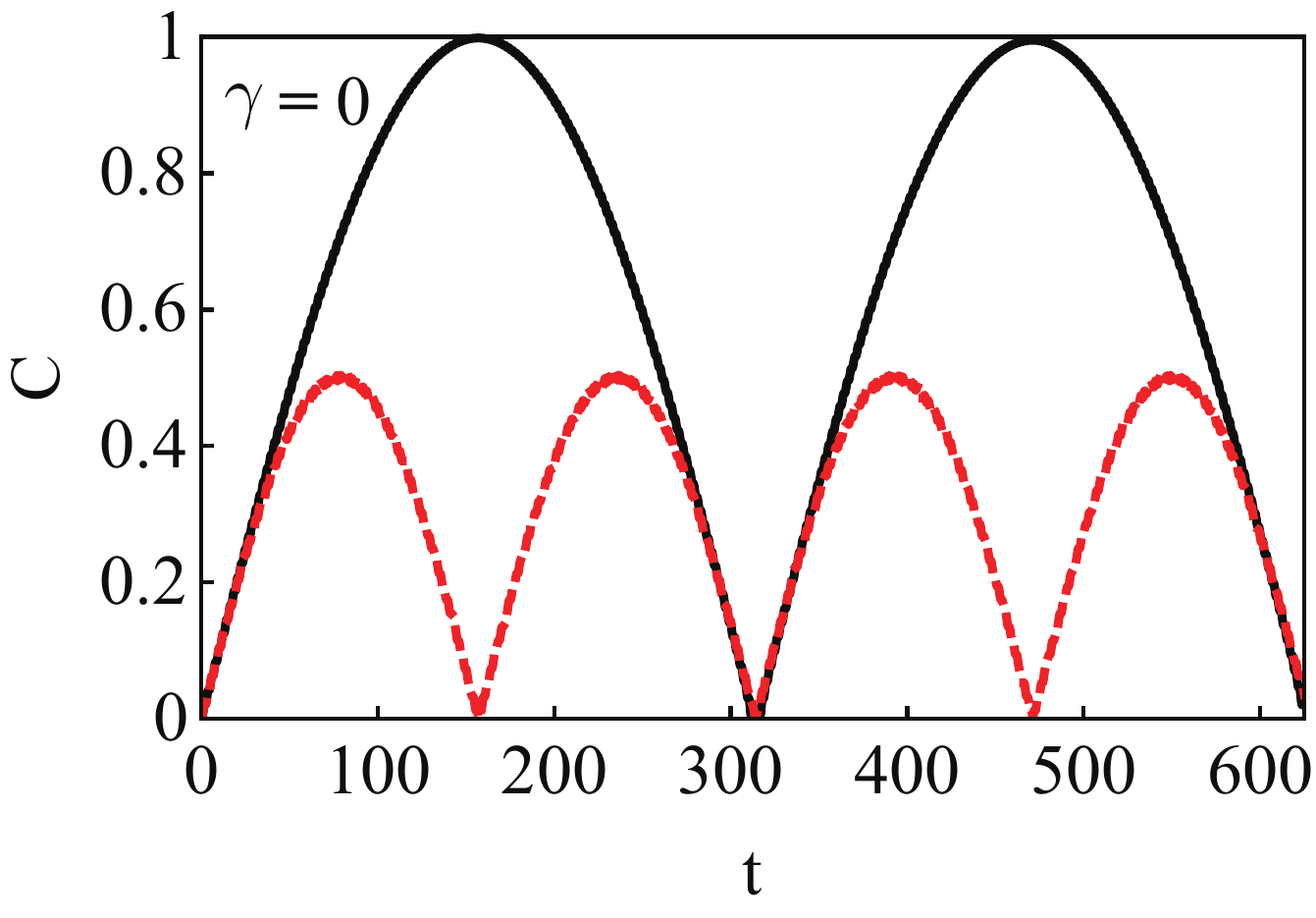}
\put(-215,135){(a)}

\includegraphics[width=0.42\textwidth]{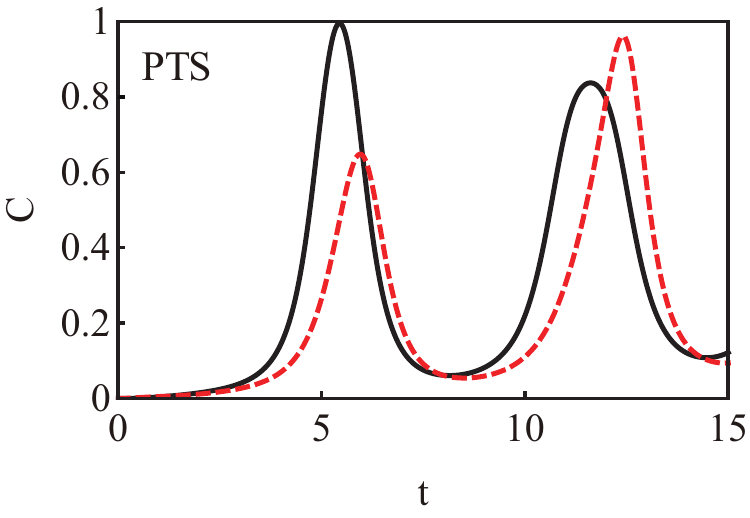}
\put(-215,135){(b)}

\caption{Dynamic evolution of entanglement described by the concurrence. (a) The Hermitian case. The parameters are taken as  $\gamma=0$, $J=0.01$, and $\Omega=1.5$. (b) The non-Hermitian case with the PT-symmetric Hamiltonian under consideration. The parameters are taken as  $\gamma=1.1$, $J=0.01$, and $\Omega=1.5$.  The solid and dashed lines correspond to the initial-state parameter  $\theta=\pi/2$ and  $\theta=\pi/4$, respectively.}
		 \label{fig8}
	\end{figure}

\section{EP quantum sensing }

In this section, we study eigenstate quantum sensing to system parameters in our model. We show that the parameter sensitivity of the quantum sensing can be regarded as the signature to witness of EPs. Quantum sensing exploits the unique characteristics of quantum phenomena to achieve ultra-precise measurements \cite{V. Giovannett,S. Lloyd,C. L. Degen}. It has become a rapidly developing research branch in the field of quantum science and technology. Recent studies have shown that the EPs of physical systems with the PT symmetry can enhance the optical sensing in the classical region \cite{H. M. Wiseman,X.-W Luo}.

Generally speaking, the sensitivity of parameter estimation is characterized by quantum Fisher information (QFI) \cite{66,67}, which can evaluate the sensitivity limit, i.e., the Cram\`{e}r-Rao bound, without reference to a specific measurement scheme. The standard deviation of an estimated value $\xi$ in a single parameter measurement is bounded by the Cram\`{e}r-Rao bound  $\delta\xi_{CR}=1/\sqrt{\mathcal{F}(\xi)}$. This means that the QFI gives the best theoretical accuracy of the parameter to be estimated \cite{S. L. Braunstein}.

We numerically calculate the QFI of eigenstates related to EPs with respect to the parameters of the system under our consideration. When the eigenstate is nondegenerate, the QFI is a smoothing function of the parameters to be measured. However, under the condition of eigenstate degeneracy, it has a sharp peak in EPs \cite{X. F. Ou}. Therefore, the QFI is a signature for the appearance of EPs. In the following, as an example, we take two parameters $J$ and $\Omega$ as estimated parameters  to calculate the quantum Fisher information of the eigenvector $|\Psi_{3}\rangle$ given  by Eq. (10). Since $|\Psi_{3}\rangle$ is a pure state, the QFI can be calculated by the following expression
\begin{align}
\mathcal{F}_\kappa=4[\langle\partial_{\kappa}\Psi_{3}|\partial_{\kappa}\Psi_{3}\rangle-|\langle\partial_{\kappa}\Psi_{3}|\Psi_{3}\rangle|^{2}].
\end{align}
where  $\kappa=J, \Omega$, and $|\partial_{\kappa}\Psi\rangle$ is the derivative with parameter $\kappa$.

\begin{figure}[t]
		\includegraphics[width=0.42\textwidth]{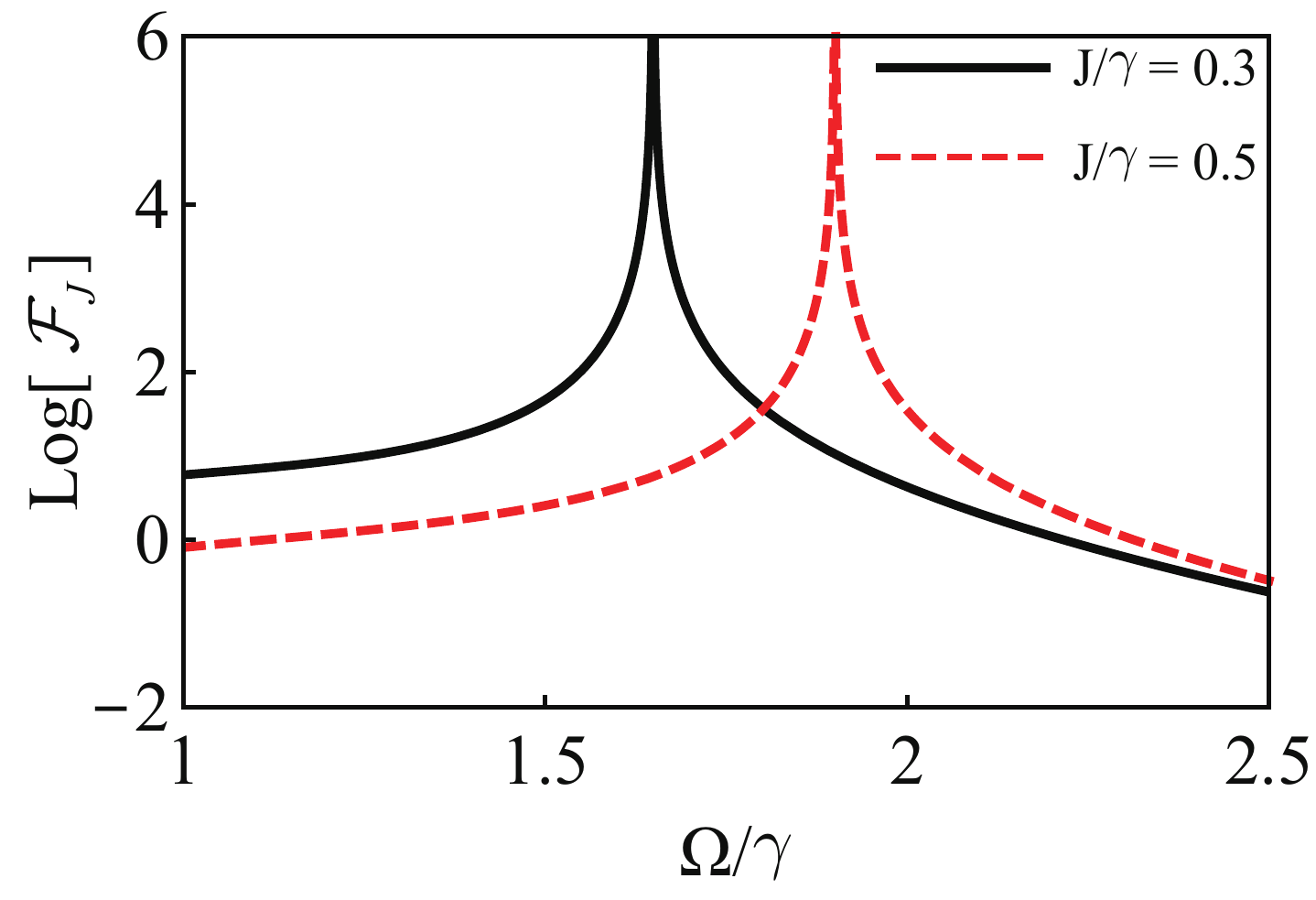}
\put(-215,135){(a)}

\includegraphics[width=0.42\textwidth]{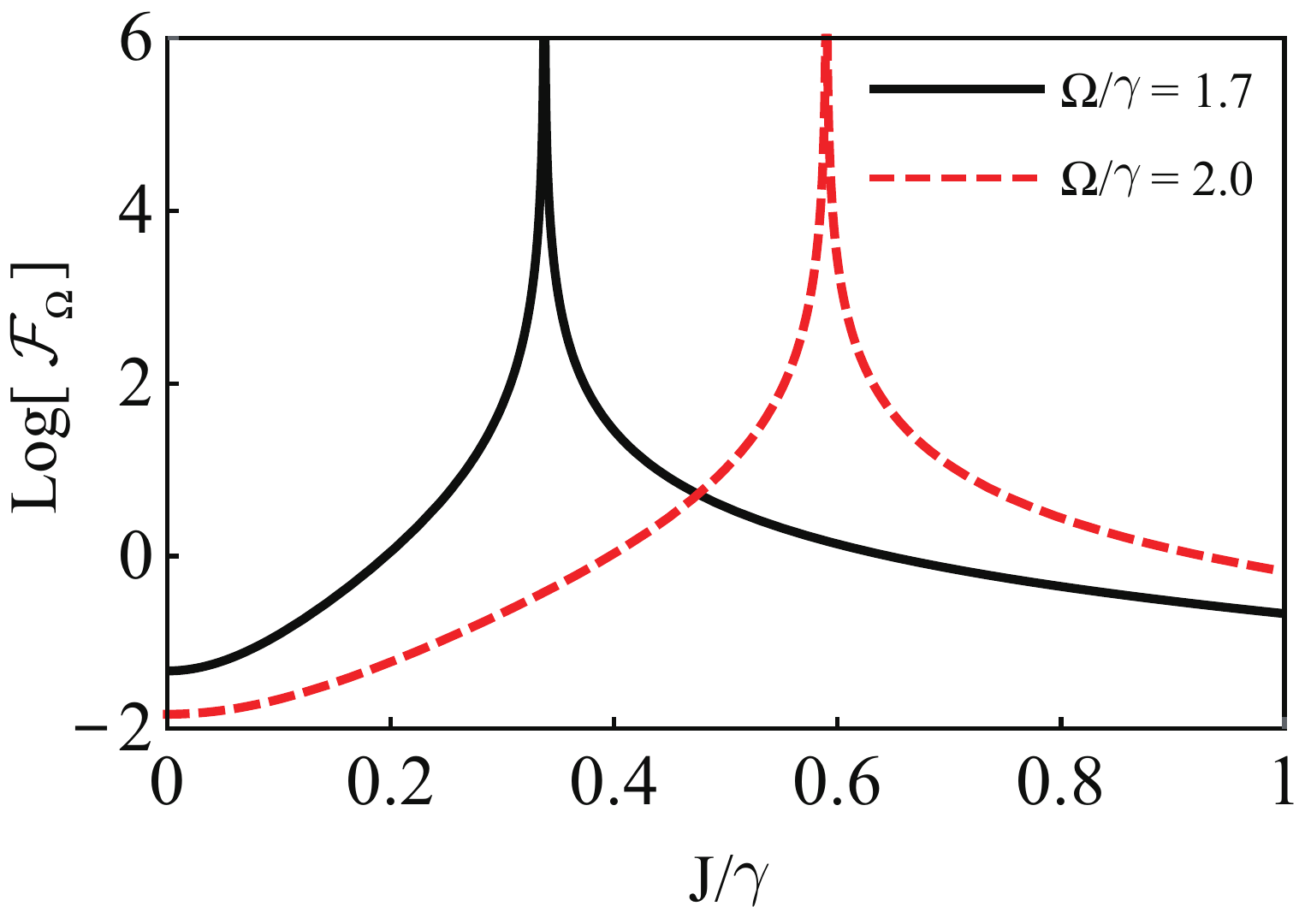}
\put(-215,135){(b)}
		\caption{(a) The logarithm of QFI via changing the  parameter $\Omega$. The black solid line (red dashed line)  parameters is $J=0.300$ ($J=0.500$), and  its critical parameters  are $\Omega_c=1.649$, $J_c=0.300$($\Omega_c=1.900$, $J_c=0.500$). (b) The logarithm of QFI via changing the inter-qubit interaction $J$ . The black solid line (red dashed line)   parameters is $\Omega=1.700$ ($\Omega=2.000$), and  its critical parameters  are $\Omega_c=1.700$, $J_c=0.338$($\Omega_c=2.000$, $J_c=0.589$). } \label{fig9}
\end{figure}

In Fig. \ref{fig9},  we plot the QFI logarithm  of the eigenvector $|\Psi_{3}\rangle$ with respect to  two parameters $J$ and $\Omega$.   Fig. \ref{fig9} (a)  shows the change of the QFI logarithm with respect to the estimated parameter $\Omega$ when the parameter $J/\gamma$ takes 0.3 and 0.5, respectively. In this case,  the two EPs appear at $(\Omega_c, J_c)=(1.649,0.300)$ and $ (1.900,0.500)$. They correspond to two peaks of the QFI, respectively.  The left-side phase of each peak is the PTB phase while the right-side phase of each peak is the PTS phase. In the PTB (PTS) phase the QFI increases (decreases) with the increase of the parameter $\Omega$.      The QFI is divergent at EPs. It is that the sudden change of the QFI at EPs can witness the phase transition between the PTB and PTS phases.
Similarly,  the PTB-PTS phase transition can also be witnessed through the QFI  with respect to the estimated parameter $J$.  Fig. \ref{fig9} (b)  shows the change of the QFI logarithm with respect to the estimated parameter $J$ when the parameter $\Omega/\gamma$ takes 1.700 and 2.00, respectively.  The two peaks of the QFI appear at EPs  $(\Omega_c, J_c)=(1.700,0.338)$ and $ (2.000,0.589)$. However, the phase regimes are different from those of Fig.  \ref{fig9} (a).  In Fig. \ref{fig9} (b), the left-side phase of each peak is the PTS phase while the right-side phase of each peak is the PTB phase. In the PTS (PTB) phase the QFI increases (decreases) with the increase of the parameter $J$.
It should be pointed out that the QFI gives the Cram\'{e}r-Rao bound of the estimated parameter $\kappa$  with $\delta\kappa_{CR}=1/\sqrt{\mathcal{F}_\kappa}$, and it is independent of a measurement scheme. So we need to look for a concrete measurement scheme to detect the existence of the EPs.


We here propose an EP-sensing scheme by measuring the quantum coherence of a single qubit. We show that the EPs-sensing scheme can detect not only  EPs but also the regimes of the PTB and PTS phases.
Quantum coherence of the first qubit can be characterized by the operator $(\sigma^{1}_{x})$. The precise sensitivity of the estimated parameter $\kappa$ is given by the error propagation formula
\begin{align}
\delta_{\kappa}^{2}=&\frac{(\triangle \sigma^{1}_{x})^{2}}{(\partial_{\kappa}\langle\sigma^{1}_{x}\rangle)^2},
\end{align}
where the variance of the operator $(\sigma^{1}_{x})$ is given by
\begin{align}
&(\triangle \sigma^{1}_{x})^{2}=1-(\langle\Psi_{3}| \sigma^{1}_{x} |\Psi_{3}\rangle)^{2}.
\end{align}

\begin{figure}[t]
		\includegraphics[width=0.42\textwidth]{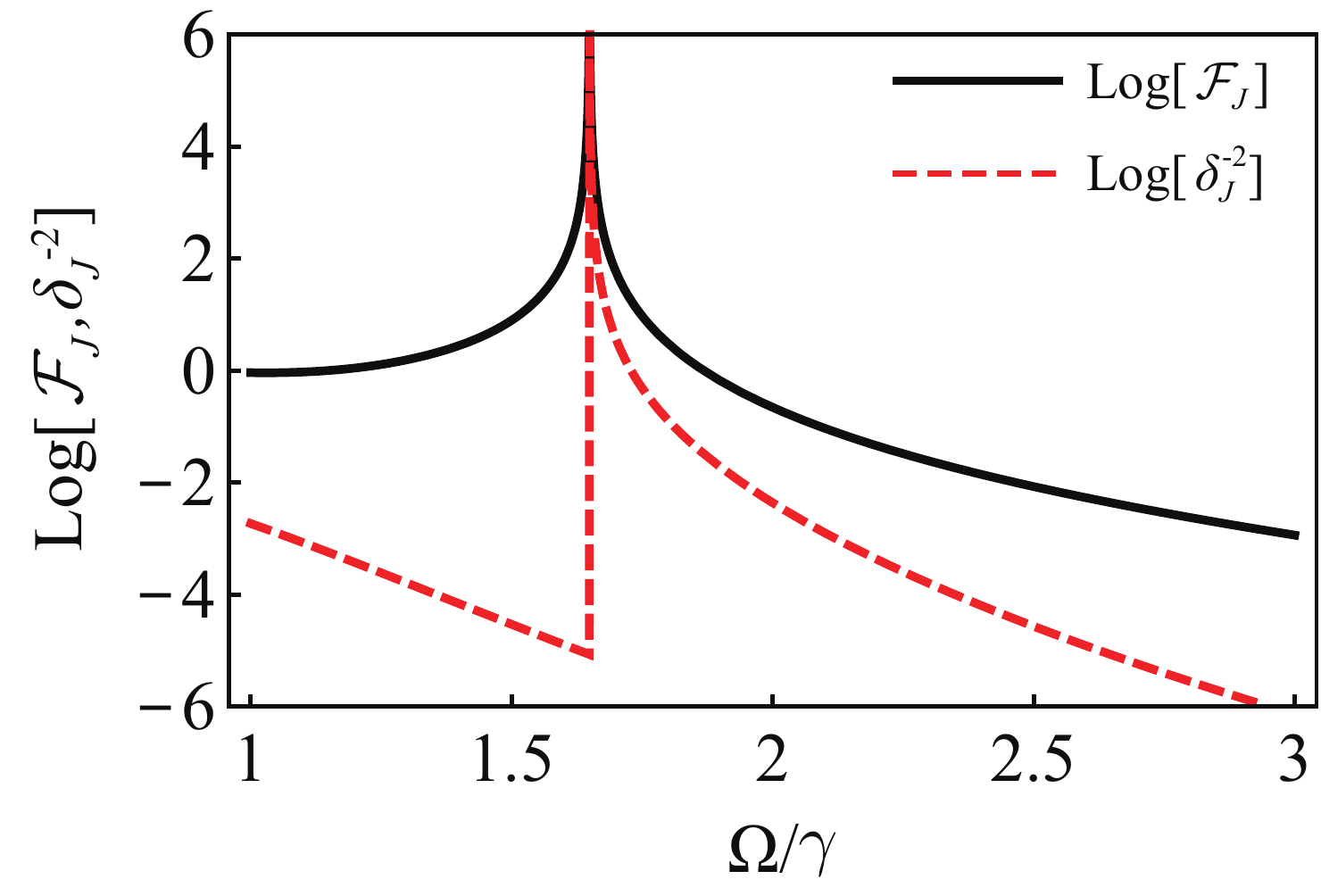}
\put(-215,135){(a)}

\includegraphics[width=0.42\textwidth]{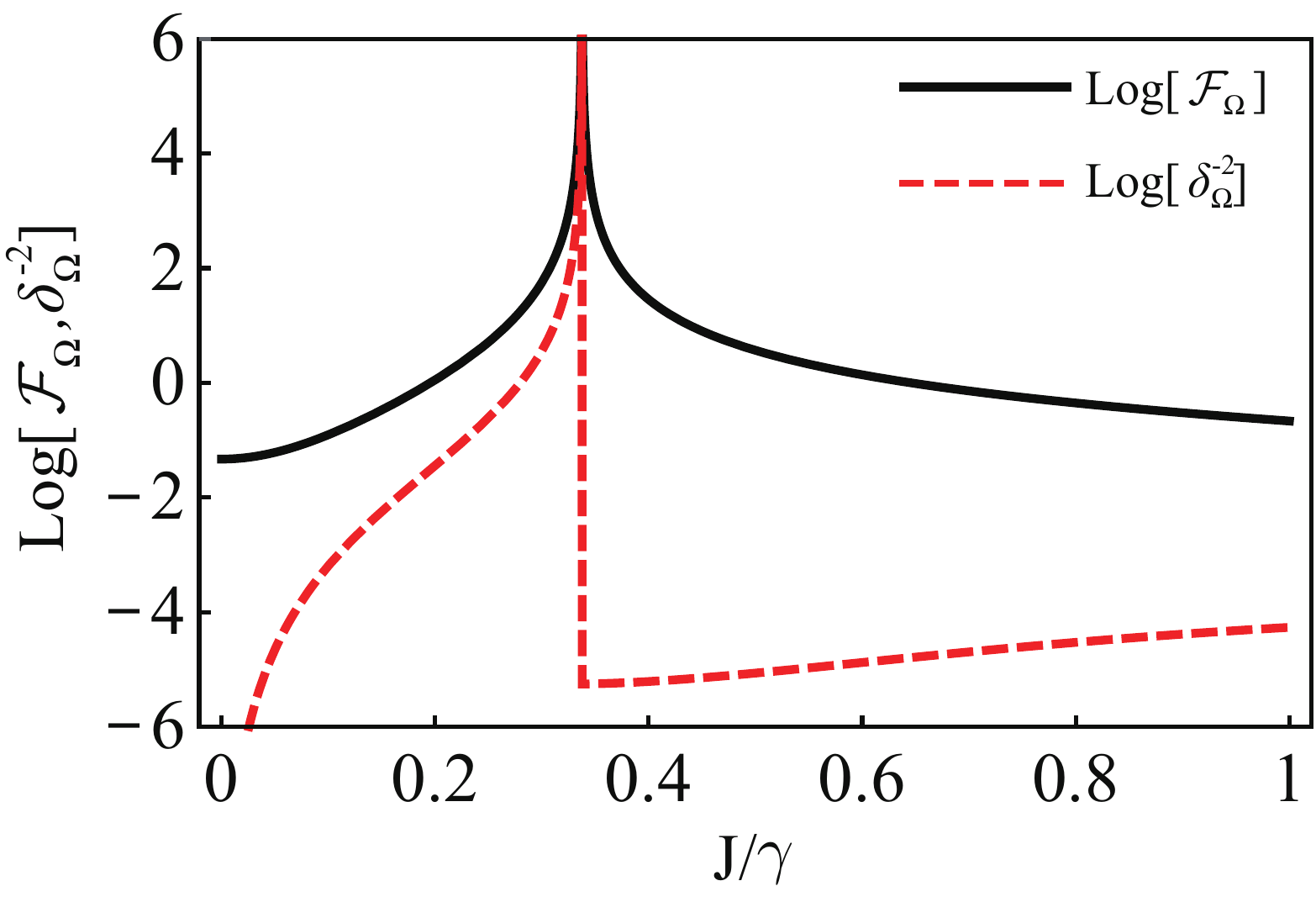}
\put(-215,135){(b)}
\caption{(a) The logarithm of inverse variance (the red dashed line) and  the logarithm of QFI (the black solid line) as functionse via changing parameter $\Omega$. The system parameter  are $J=0.300$, and  its critical parameters  are $\Omega_c=1.649$, $J_c=0.300$. (b)The logarithm of inverse variance (the red dashed line) and  the logarithm of QFI (the black solid line ) as functionse via changing the inter-qubit interaction $J$. The system parameter is  $\Omega_c=1.700$ and  its critical parameters  are $\Omega_c=1.700$, $J_c=0.338$. } \label{fig10}
\end{figure}

In Fig. 8, we plot the variances of measurement parameters $\kappa$ ($\kappa=J, \Omega)$ (the dashed lines). In order to compare them with the QFI, we also plot their corresponding QFI (the solid lines).
The dashed line in Fig. 8 describes the inverse of the variance squares ($\delta^{-2}_{\kappa}$  for the estimated parameter $\kappa$.  From Fig. 8 we can see both $\log\delta^{-2}_{\kappa}$ and $\log \mathcal{F}_{\kappa}$ are divergent at the same point which is just the EP of the system under our consideration. The  parameter values of the EPs in Fig. 8(a) and 8(b) are  $(\Omega_c, J_c)=(1.649,0.300)$  and $(1.700,0.338)$, respectively.   Hence, we can detect EPs through measuring the quantum coherence of the single qubit.

It is interesting to note that the PTS (PTB) phase in Fig. 8(a) and the PTS (PTB) phase in Fig. 8(b)  are on different sides of the EPs.
In Fig. 8(a) the PTS (PTB) phase is on the right (left) side of the EP while in Fig. 8(b) the PTS (PTB) phase is on the left (right) side of the EP.
In the PTS phase, the higher is the precise sensitivity of the estimated parameter, more near the EP of the manipulating parameters.  At EPs, the precise sensitivity can attain the Cram\'{e}r-Rao bound. This represents a general feature for those proposals that utilize criticality and the associated divergent behaviors for sensing. Therefore,  the achievable measurement precision can be significantly enhanced near the EPs in the PTS phase.

\begin{figure}[t]
\includegraphics[width=0.36\textwidth]{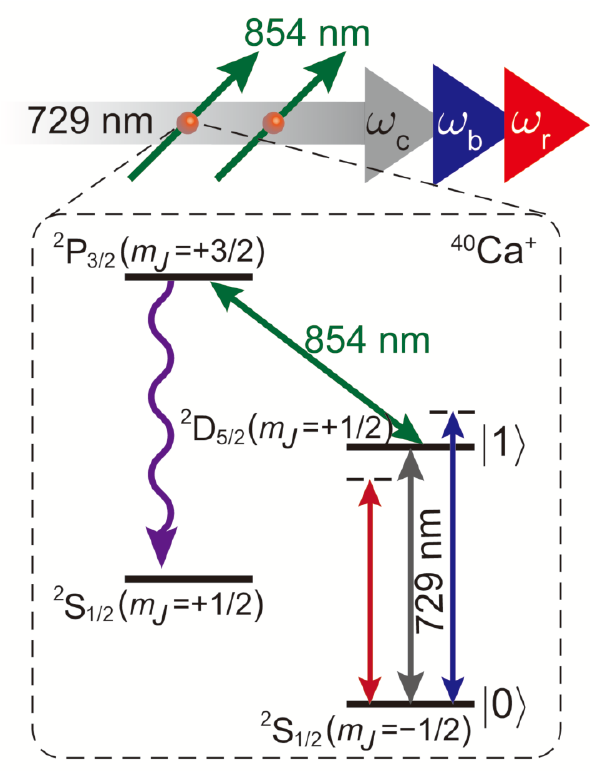}
\caption{Realization of PT symmetric two-qubit system in a linear trap containing two individually addressed $^{40}\mathrm{Ca}^+$ ions. Three 729-nm lasers and one 854-mn laser are applied to the ions. 729-nm laser with frequency $\omega_c$ creates the coherent Rabi coupling $\Omega$ between the states $|\downarrow\rangle \leftrightarrow |\uparrow\rangle$, which correspond to the energy levels $^2S_{1/2}(m_J=-1/2)$ and $^2D_{5/2}(m_J=+1/2)$, respectively. The other two 729-nm lasers with frequency $\omega_b$ and $\omega_r$ are applied to the ions in order to drive the blue- and red- sideband transitions with apposite detunings. This can give rise to the effective spin-spin coupling $J$. Effective decay rate $\gamma$ in state $|\uparrow \rangle$ is created by 854-nm laser, which couples $|\uparrow\rangle$ to a short-life level $^2P_{3/2}(m_J=3/2)$(it decays quickly to the state $^2S_{1/2}(m_J=1/2)$).} \label{fig11}
\end{figure}

\section{Experimental feasibility}

In order to demonstrate our theoretical model about the PT symmetric two-qubit system, we now propose an experimentally accessible scheme by using  a trapped-ion system, which is schematically shown in Fig. \ref{fig11}. Effective two-level spins are represented by energy levels $|^2S_{1/2}(m_J=-1/2)\rangle$ and $|^2D_{5/2}(m_J=+1/2)\rangle$ of $^{40}\mathrm{Ca}^+$ ions with long coherence time exceeding 1s \cite{W. C. Wang}. The coherent Rabi coupling is created by a 729-nm laser with frequency $\omega_c$. The other two 729-nm lasers with frequency $\omega_b$ and $\omega_r$ are applied to the ions in order to drive the blue- and red- sideband transitions with apposite detunings. This can give rise spin-dependent forces and excite virtual phonons which mediate required interacting Hamiltonian $J\sigma_z^1\sigma_z^2$ between two spins. Another laser at 854-nm induces a tunable decay rate $\gamma$ in state $|\uparrow \rangle$, by coupling $|\uparrow\rangle$ to a short-life level $|^2P_{3/2}(m_J=3/2)\rangle$ which decays quickly to the state $|^2S_{1/2}(m_J=1/2)\rangle$. The effective two-spin system is well described by the non-Hermitian Hamiltonian $H_{eff}= H - i\gamma \mathbf{I}$, where $H$ is the PT-symmetric Hamiltonian shown in Eq. (1) with balanced gain an loss. Therefore, the corresponding matrix state of the PT symmetric two-qubit system can be get by $\rho_{PT} = e^{2\gamma t} \rho_{eff}$, which is also known as passive PT symmetry system realized in many experiments  \cite{Y. Wu,M. Naghiloo,W. C. Wang, Li2019}.

\section{Conculsions}
We have studied EP effects and eigenstate quantum sensing in the PT-symmetric two-qubit system with the Ising-type coupling. We have shown that EP properties of the system under consideration can be characterized by analyzing the degeneracy of energy eigenvalues or entanglement of eigenstates. We have investigated the entanglement dynamics of the two qubits. We have found that  the PT-symmetric two-qubit
the system has quite different dynamic behaviors of entanglement in the PTS and PTB phases.  We have revealed two remarkable dynamic EP effects during the long-time evolution, i.e.,  the steady-state entanglement in the PT-broken phase and the collapse-revival phenomenon of entanglement in the PT-symmetric phase. The two EP effects can be regarded as the signature of the PTS and PTB phases. On the other hand,  we have indicated that the PT-symmetric system can more rapidly create entanglement than a corresponding Hermitian system in the short-time-evolution region. Moreover, We have studied eigenstate quantum sensing of the PT-symmetric system and found that  the sensitivity of quantum sensing for the parameters exhibits the sudden enhancement at EPs  by calculating the QFI. We have shown that quantum-coherence measurement for the single qubit can be used to witness the existence of EPs. It could be expected that EP-enhanced quantum sensing has powerful applications in weak signal detection and precise measurement of physical parameters by tuning close to the critical point. Despite its simplicity, the current model considerably extends the phenomenology of nonequilibrium dynamics beyond that commonly assumed for dissipative systems.

\section*{Acknowledgments}
We thank Professor Jing Liu and Jiahao Huang  for useful discussions. Y.L.Z is supported by the Natural Science Foundation of Hunan Province of China (Grant No. 2023JJ30626). P.X.C. is supported by  the NSFC (Grants No 12074433). H.J. is supported by the NSFC (Grant nos. 11935006 and 11774086) and the Science and Technology Innovation Program of Hunan Province (Grant no. 2020RC4047). L.M.K. is supported by the NSFC (Grant Nos. 12247105, 1217050862, and 11935006).

\appendix

\section{Exceptional-point analysis of Haniltonian (1)}

In this appendix we present an exceptional-point analysis of Hamiltonian (1). We show that Hamiltonian (1) has second-order EPs associated with egienvalues $E_3$ and $E_4$.  Let $Y=r e^{i\theta}$. From Eqs. (3)-(7) we can rewrite $E_3$ and $E_4$ as
\begin{equation}
 E_{3}=\frac{1}{3}(J-e_3), \hspace{0.3cm} E_{4}=\frac{1}{3}(J-e_4),
\end{equation}
where  we have introduced
\begin{eqnarray}
e_{3}&=&\frac{X}{r}e^{-i\left(\theta-\frac{\pi}{3}\right)} + r e^{i\left(\theta-\frac{\pi}{3}\right)}, \\
e_{4}&=&\frac{X}{r}e^{-i\left(\theta+\frac{\pi}{3}\right)} + r e^{i\left(\theta+\frac{\pi}{3}\right)},
\end{eqnarray}
where $X$ is a real number.

From Eq. (A1)-(A3), we can obtain the real and imaginary parts of  $E_3$ and $E_4$ with the following expressions
\begin{eqnarray}
3\text{Re}E_{3}&=&J-\left(\frac{X}{r}+r\right) \cos\left(\theta-\frac{\pi}{3}\right), \\
3\text{Im}E_{3}&=&-\left(\frac{X}{r}-r\right) \sin\left(\theta-\frac{\pi}{3}\right), \\
3\text{Re}E_{4}&=&J-\left(\frac{X}{r}+r\right) \cos\left(\theta+\frac{\pi}{3}\right), \\
3\text{Im}E_{4}&=&-\left(\frac{X}{r}-r\right) \sin\left(\theta+\frac{\pi}{3}\right).
\end{eqnarray}

From Eqs. (A4)-(A7)	we can find that the critical  parameters $J_c$ and $\Omega_c$ at EPs must obey the following equations
\begin{equation}
\theta(J_c, \Omega_c)=0,  \hspace{0.3cm} X(J_c, \Omega_c)=r^2(J_c, \Omega_c).
\end{equation}
At above EPs we have
\begin{eqnarray}
\text{Re}E_{3}&=&\text{Re}E_{4}=\frac{1}{3}\left[J-\frac{1}{2}\left(\frac{X}{r}+r \right)\right], \\
\text{Im}E_{3}&=&\text{Im}E_{4} =0.
\end{eqnarray}
which indicates that the two energy eigenvalues become real and degenerate. Substituting the condition of the critical parameters (A8) into Eqs. (10)-(14). It is easy to find that
\begin{eqnarray}
|\Psi^{R}_{4}\rangle&=&|\Psi^{R}_{3}\rangle  \nonumber\\
&=&N(c_0|0,0\rangle +c_1|0,1\rangle +c_1|1,0\rangle +|1,1\rangle). \nonumber\\
\end{eqnarray}
where $N$ is the normalization constant and the two superposition coefficients $c_0$ and $c_0$ are given by
\begin{eqnarray}
c_0 &=&-2(J+E)(J-E +i\gamma)/\Omega^{2}-1, \\
c_1&=&-(J-E +i\gamma)/\Omega, \\
E&=&\frac{1}{3}\left[J-\frac{1}{2}\left(\frac{X}{r}+r \right)\right].
\end{eqnarray}

Therefore, we demonstrate  the coalescence of eigenvalues $E_3$ and $E_4$  and their corresponding  eigenstates of the non-Hermitian Hamiltonian with PT symmetry given by Eq. (1) at EPs.


\end{document}